\documentclass[twocolumn,showpacs,epsfig,prl]{revtex4}
\usepackage{graphicx}
\usepackage{amssymb}
\usepackage{color,soul}

\newcommand{\ket}[1]{\left|#1\right>}
\newcommand{\bra}[1]{\left<#1\right|}

\newcommand{\mr}{\mathrm}

\newcommand{\tr}{\mathrm{tr}}
\begin{document}

\title{Heat engine driven by purely quantum information}
\author{Jung Jun Park$^{1}$}
\author{Kang-Hwan Kim$^{2}$}
\author{Takahiro Sagawa$^{3,4}$}
\author{Sang Wook Kim$^{5}$}
%\email{hyok07@gmail.com}
\affiliation{$^{1}$Department of Physics, Pusan National University, Busan 609-735, Korea\\
$^{2}$Department of Physics, Korea Advanced Institute of Science and Technology, Daejeon 305-701, Korea\\
$^{3}$The Hakubi Center for Advanced Research, Kyoto University,
%Yoshida-ushinomiya cho, Sakyo-ku,
Kyoto 606-8302, Japan\\
$^{4}$Yukawa Institute for Theoretical Physics, Kyoto University,
%Kitashirakawa-oiwake cho, Sakyo-ku,
Kyoto 606-8502, Japan\\
$^{5}$Department of Physics Education, Pusan National University, Busan 609-735, Korea
}
\date{\today}

\begin{abstract}
The key question of this paper is whether work can be extracted from a heat engine by using purely quantum mechanical information. If the answer is yes, what is its mathematical formula? First, by using a bipartite memory we show that the work extractable from a heat engine is bounded not only by the free energy change and the sum of the entropy change of an individual memory but also by the change of quantum mutual information contained inside the memory.
We then find that the engine can be driven by purely quantum information, expressed as the so-called quantum discord, forming a part of the quantum mutual information. To confirm it, as a physical example we present the Szilard engine containing a diatomic molecule with a semi-permeable wall.
\end{abstract}

% 03.67.-a Quantum information
% 89.70.Cf Entropy and other measures of information
% 05.70.-a Thermodynamics

% 05.30.-d Quantum statistical mechanics
% 05.70.Ln Nonequilibrium and irreversible thermodynamics
% 03.65.Ta Foundations of quantum mechanics; measurement theory

\maketitle

Maxwell first recognized the subtle role of information in thermodynamics, and devised his famous demon who might violate the second law of thermodynamics \cite{Leff03}. Szilard then proposed a simple physical model to realize Maxwell's demon, and claimed that information should play a role of physical entropy unless the second law is wrong \cite{Szilard29}. Now it is widely accepted that the so-called Szilard engine (SZE) does not violate the second law. The measurement process or the erasure of demon's memory saves the second law \cite{Brillouin51,Landauer61,Bennett82,Maruyama09}. The SZE indeed demonstrates how information is exploited to extract physical work, so that it may be called an information heat engine (IHE). Such an IHE has been realized in experiment \cite{Toyabe10}. One might ask ``{\em What} information is exploited?'' The correlation between an engine and a demon's memory should be responsible to it since work is extracted from the feedback control based upon the measurement outcome obtained by the demon. Note that we use the memory with the same meaning as the demon. More precisely the memory represents physical realization of rather vague terminology, the demon.

It has been proposed that the extractable work is given by the so-called QC mutual information between an engine and a memory \cite{Sagawa08}. Here `QC' emphasizes that the {\em local} measurement (thus giving rise to classical information) on the memory is performed over the quantum composite system consisting of the engine and the memory. The information obtained from the memory should be classical since it is used for feedback control of the engine which is classical in nature. In the SZE, a particle can exist either on the right or on the left side, which is nothing but a one bit classical information. The QC mutual information is bounded by the Shannon entropy of the memory which is the maximum classical information that the memory can possess. One might ask whether {\em quantum} information can be used in an IHE. If yes, what is the mathematical expression of the work from it? Even though The {\em quantum} SZE has been studied \cite{SW_Kim11}, only its dynamics is treated quantum mechanically while the information exploited is still classical \cite{KH_Kim11}.

\begin{figure}
  \includegraphics[width=9cm]{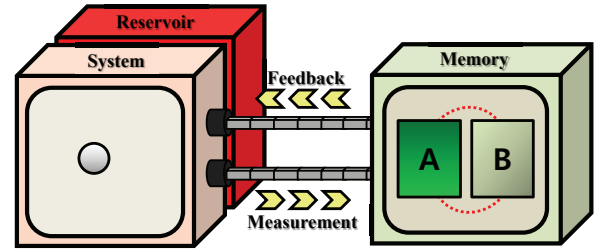}
  \caption{The schematic picture showing the setup considered here. The system is attached to the reservoir, and is measured and controlled by the memory consisting of A and B. (See the text for the detail.) }
  \label{fig1}
\end{figure}

There have been several works on the IHE using quantum information or entanglement. The entanglement initially forms between an engine and a memory \cite{Zurek03}, between a system and an observer \cite{Rio11}, and between an engine A and B when the engine consists of two parts \cite{Funo12}. The work extracted only from quantum information (correlation) is expressed as the discord \cite{Ollivier02}, the deficit \cite{Oppenheim02}, or the conditional von Neumann entropy \cite{Rio11}. However, the initial state of a heat engine or a system should be in thermal equilibrium, namely satisfies the canonical distribution with the well defined temperature if the thermodynamic work is extracted from them. It is wonder how quantum entanglement survives when the system contacts with the heat reservoir so as to be in equilibrium.

In this Letter we show (i) the quantum mutual information of {\em correlated} memories of the IHE can be used to generate work, (ii) the amount of work extracted from purely quantum mechanical information is expressed as the discord, and (iii) present a realizable physical model, namely a SZE containing a diatomic molecule with a semi-permeable wall \cite{Neumann55}. Our results are general in that we consider a full quantum composite system consisting of all possible physical components such as an engine, a reservoir, and a memory. We emphasize that the quantum entanglement or correlation exists only in the memory.

We consider a thermodynamic process of a system $S$, e.g. the SZE, which is assumed to have two states or one bit, e.g. the left and the right side of the SZE, and interact with a heat reservoir $R$ at temperature $T$ \cite{SW_Kim11}. A demon $M$ consists of two one-bit memories, $A$ and $B$. It is noted that each memory is indeed not necessarily one bit in our theory, but only for simplicity here we assume it has one bit. Even if one bit memory, namely only $A$, is enough to describe $S$ of two states, we intentionally introduce the second, namely $B$ to investigate the role of quantum entanglement in the IHE. Two bit memories can be realized by atomic internal states. It is known that which-way information of the center of mass of a two-level atom in a matter wave double slit experiment can be encoded into its internal states \cite{Durr98}, which are equivalent to an one bit memory. We then regard the two internal states, namely $A$ and $B$, of a heteronuclear diatomic molecule as two bit memories. {\em These states are not necessarily in thermal equilibrium so that they can form quantum entangled states.} Note that two atoms with entangled internal states has been considered in the context of photon Carnot engine \cite{Dillenschneider09}.

The total Hamiltonian is written as
\begin{equation}
H(t) = H_{SR}(t) + H_{SM_{AB}}^{int}(t) + H_{M_{AB}}(t),
\end{equation}
where $H_{SR}(t)$, which reads $H_{SR}(t) = H_S(t) + H_R + H_{SR}^{int}(t)$, describes the system, the reservoir, and their interaction, respectively. $H_{SM_{AB}}^{int}(t)$ is the interaction Hamiltonian describing measurement process done by the demon. $H_S(t)$ and $H_{M_{AB}}(t)$ are the Hamiltonian of $S$ and the memory $AB$, respectively, which are controlled by varying external parameters such as an applied magnetic field or volume of the gas. The thermodynamic process of the total system is divided into four stages.

\textit{Stage 0 (Initial state)} - The system $S$ contacts with the reservoir $R$ at temperature $T$ so that it is in thermodynamic equilibrium. The density matrix of the initial state of the total system reads
\begin{equation}
\rho^{(i)} = \rho_{AB}^{(i)} \otimes \rho_{SR}^{(i)},
\end{equation}
with
\begin{equation}
\rho_{SR}^{(i)} = \frac{\mr{exp}(-\beta H_S^{(i)})}{Z_S^{(i)}} \otimes \frac{\mr{exp}(-\beta H_R)}{Z_R},
\label{eq_rho_SR_i}
\end{equation}
where $\beta = (k_B T)^{-1}$, $H_S^{(i)}=H_S(0)$, $Z_S^{(i)} = \tr\lbrace\mr{exp}(-\beta H_S^{(i)})\rbrace$, and $Z_R = \tr\lbrace\mr{exp}(-\beta H_R)\rbrace$. Note that there is no restriction on $\rho_{AB}^{(i)}$.

\textit{Stage 1 (Unitary evolution)} - Before measurement we perform a thermodynamic process on the system and the reservoir with the memory intact. The state is then transformed to
\begin{equation}
\rho^{(1)} = U^{(1)}\rho^{(i)} {U^{(1)}}^\dagger
\end{equation}
with $U^{(1)} = I_{AB} \otimes U_{SR}^{(1)}$. Through this paper $I_X$ denotes the identity operator for $X$. In the SZE, inserting a wall corresponds to this stage.

\textit{Stage 2 (POVM)} - The measurement of $S$ is done by using positive operator valued measures (POVMs) \cite{Nielsen-Chuang00}. In order to study the role of the quantum correlation of the IHE, the measurement is performed only by $A$ with the rank-1 projector $\Pi_A^k$ for obtaining each outcome $k$ with the probability $p_k$, instead of the whole $AB$. The density matrix after the measurement is given as
\begin{equation}
\rho^{(2)} = \sum_k \Pi_A^k {\rho^{(1)}}' \Pi_A^k = \sum_k p_k \ket{k}_A\bra{k} \otimes \rho_{BSR}^{(2)k},
\end{equation}
where ${\rho^{(1)}}' = U^{(2)}\rho^{(1)} {U^{(2)}}^\dagger$ with $U^{(2)}$, an unitary operator generating correlation between $S$ and $A$, $p_k = \tr\left[\Pi_A^k {\rho^{(1)}}' \Pi_A^k\right]$, and $\rho_{BSR}^{(2)k} = \tr_A\left[\Pi_A^k {\rho^{(1)}}' \Pi_A^k / p_k\right]$.

If the measurement is performed by the whole $AB$, namely $M$, one instead obtains
\begin{equation}
\sigma^{(2)} = \sum_k \Pi_M^k {\sigma^{(1)}}' \Pi_M^k = \sum_k q_k \ket{k}_M\bra{k} \otimes \sigma_{SR}^{(2)k},
\end{equation}
where ${\sigma^{(1)}}' = V^{(2)}\rho^{(1)} {V^{(2)}}^\dagger$ with $V^{(2)}$, an unitary operator generating correlation between $S$ and $M$, $q_k = \tr\left[\Pi_M^k {\sigma^{(1)}}' \Pi_M^k\right]$, and $\sigma_{SR}^{(2)k} = \tr_M\left[\Pi_M^k {\sigma^{(1)}}' \Pi_M^k / q_k\right]$. As mentioned earlier, it has been  shown that the work bound of this IHE is given by the QC mutual information between $S$ and $M$, namely $W_{ext} \leq -\Delta F_S +k_B T I_{QC}(S:X)$ \cite{Sagawa08}, where $F_S$ and $X$ denotes the Helmholtz free energy of $S$ and the set of outcomes $k$'s, respectively.
Here $I_{QC}(S:X)$ is defined as $S(\rho_S^{(i)})-\sum_k q_k S(\sigma_{S}^{(2)k})$, where $\rho_S^{(i)}=\rm{tr}_R \left[\rho_{SR}^{(i)}\right]$ and $\sigma_{S}^{(2)k}=\rm{tr}_R \left[\sigma_{SR}^{(2)k}\right]$.

\textit{Stage 3 (Feedback control)} - Formally the feedback control can be described by a unitary transform of the total system, namely $U^{(3)} = I_B \otimes  \sum_{k} \ket{k}_A\bra{k} \otimes U^{k}_{SR}$. The final state at $t=t_f$ reads
\begin{equation}
\rho^{(f)} = U^{(3)} \rho^{(2)} {U^{(3)}}^{\dagger}.
\end{equation}
It is noted that the final state is not necessarily the canonical distribution \cite{Sagawa08}, but this makes no problem below [See Eq.~(\ref{eq_entropy difference_if_2_klein})].

Now let us find the entropy change of $SR$ during the above mentioned thermodynamic process. Note that $S(\rho) = -\tr(\rho \mr{ln} \rho)$ represents the von Neumann entropy and $H(p_k) = -\sum_k p_k \mr{ln} p_k$ the Shanon information.
Since the measurement performed in the stage 2 increases the entropy, i.e. $S[\rho^{(i)}] \leq S[\rho^{(2)}]$, one obtains
\begin{equation}
S[\rho_{SR}^{(i)}] + S[\rho_{AB}^{(i)}] \leq H(p_k) + \sum_k p_k S[\rho_{BSR}^{(2)k}].
\label{eq_entropy_total_if}
\end{equation}
Due to the subadditivity of von Neumann entropy, Eq.~(\ref{eq_entropy_total_if}) is rewritten as
\begin{equation}
S[\rho_{SR}^{(i)}] - \sum_k p_k S[\rho_{SR}^{(2)k}] \leq H(p_k) + \sum_k p_k S[\rho_B^{(2)k}] - S[\rho_{AB}^{(i)}].
\label{eq_entropy_total_if_2}
\end{equation}
Considering the concavity of the Neumann entropy, we obtain after some algebra
\begin{equation}
\label{eq_entropy difference_if_1}
S[\rho_{SR}^{(i)}] - S[\rho_{SR}^{(f)}] \leq \Delta S_A + \Delta S_B - \Delta I,
\end{equation}
where $\Delta S_{X} = S[\rho_{X}^{(f)}] - S[\rho_{X}^{(i)}]$ with $X \in \{A,B\}$ and $\Delta I = I(A^{(2)}:B^{(2)}) - I(A^{(i)}:B^{(i)})$. Here $I$ denotes the quantum mutual information, $I(A:B) = I(B:A) = S(\rho_A) + S(\rho_B) - S(\rho_{AB})$.

Next, let us find the bound of the work extractable from this IHE. By using Klein's inequality and Eq.~(\ref{eq_entropy difference_if_1}) one obtains
\begin{equation}
S[\rho_{SR}^{(i)}] - \tr[\rho_{SR}^{(f)} \ln \rho_{SR}^{(f)can}] \leq \Delta S - \Delta I,
\label{eq_entropy difference_if_2_klein}
\end{equation}
with $\Delta S \equiv \Delta S_A + \Delta S_B$, and
\begin{equation}
\rho_{SR}^{(f)can} = \frac{\mr{exp}(-\beta H_S^{(f)})}{Z_S^{(f)}} \otimes \frac{\mr{exp}(-\beta H_R)}{Z_R},
\label{eq_rho_SR_can}
\end{equation}
where $Z_S^{(f)} = \tr\{\mr{exp}(-\beta H_S^{(f)})\}$ with $H_S^{(f)}=H_S(t_f)$.
By inserting $\rho_{SR}^{(i)}$ of Eq.~(\ref{eq_rho_SR_i}), $\rho_{SR}^{(f)}$, and $\rho_{SR}^{(f)can}$ of Eq.~(\ref{eq_rho_SR_can}) into Eq.~(\ref{eq_entropy difference_if_2_klein}) we obtain
\begin{equation}
E_S^{(i)} - E_S^{(f)} + E_R^{(i)} - E_R^{(f)} \leq F_S^{(i)} - F_S^{(f)} + k_BT[\Delta S - \Delta I]
\end{equation}
with $E_S^{(i)} = \tr[\rho^{(i)} H_S^{(i)}] $, $E_S^{(f)} = \tr[\rho^{(f)} H_S^{(f)}]$, $E_R^{(i)} = \tr[\rho^{(i)} H_R^{(i)}]$, $E_R^{(f)} = \tr[\rho^{(f)} H_R^{(f)}]$, $F_S^{(i)} = -k_BT\mr{ln}Z_S^{(i)}$, and $F_S^{(f)} = -k_BT\mr{ln}Z_S^{(f)can}$.

Because the work extractable from the engine is defined as $W_{ext} = -\Delta U_S + Q$, where $\Delta U_S = E_S^{(f)} - E_S^{(i)}$ is the change of the internal energy and $Q = E_R^i - E_R^f$ the heat exchange between $S$ and $R$, we finally reach
\begin{equation}
W_{ext} \leq -\Delta F_S + k_BT\Delta S - k_BT\Delta I
\label{eq_work_ext}
\end{equation}
with $\Delta F_S = F_S^{(f)} - F_S^{(i)}$. Here $\Delta F_S$ and $\Delta S$ describe the free energy difference and the entropy change of each memory, respectively, so that they play the role of thermodynamic entropy of a usual IHE consisting of the memory with no correlation. $\Delta I$ represents the change of the quantum mutual information or the total correlation between the memory A and B. Note that the increase of the entropy of each memory but the decrease of the correlation are exploited to generate work, which is reflected in the different signs of them in Eq.~(\ref{eq_work_ext}).

The correlation $\tilde{J}$ between $A$ and $B$ formally satisfies $\tilde{J}(B:A) = S(B) - S(B|A)$, where $S(B|A)$ represents the conditional entropy. In quantum mechanics the conditional entropy can be well defined only if the projectors of the measurement on $A$, $\{\Pi_A^i\}$, are given. Therefore, it should be written as $\tilde{J}(B:A)=S(\rho_B)-S(\rho_B|\{\Pi_A^i\})$, which obviously depends on $\{\Pi_A^i\}$. Interestingly $I(B:A)-\tilde{J}(B:A)$ does not vanish. We can thus define the quantum discord as $\delta(B|A) = \mr{min} [ I(B:A)-\tilde{J}(B:A) ]$ \cite{Ollivier02}, which also reads
\begin{equation}
\delta(B|A) = S(\rho_A) - S(\rho_{AB}) + \mr{min}\sum_i p_i S(\rho_B^i)
\label{eq_discord}
\end{equation}
with $p_k = \tr\lbrace\Pi_A^k \rho_{AB} \Pi_A^k\rbrace$, and $\rho_B^k = \Pi_A^k \rho_{AB} \Pi_A^k / p_k$. Here the minimization is performed over the sets of the projectors $\{\Pi_A^k\}$. This measures the quantum mechanical contribution of the correlation between $A$ and $B$. Thus, the quantum mutual information is reexpressed as $I(A:B) = J(B:A) +\delta(B|A)$ with $J(B:A) = \mr{max}[\tilde{J}(B:A)] = S(\rho_B) - \mr{min}\sum_k p_k S(\rho_B^k)$.

Since $\rho_{AB}^{(2)}$ is the post-measurement density matrix, one finds $I(A^{(2)}:B^{(2)}) = J(A^{(2)}:B^{(2)})$ implying $\rho_{AB}^{(2)}$ has no quantum mechanical correlation in the context of the quantum discord. It means that $\Delta \delta(B|A) = \delta(B^{(f)}|A^{(f)}) - \delta(B^{(i)}|A^{(i)}) = - \delta(B^{(i)}|A^{(i)})$ due to $ \delta(B^{(f)}|A^{(f)})=0$. Thus, Eq.~(\ref{eq_work_ext}) is rewritten as
\begin{equation}
W_{ext} \leq -\Delta F_S + k_BT\Delta S - k_BT\Delta J +k_BT \delta(B^{(i)}|A^{(i)}),
\label{eq_entropy difference_if_2}
\end{equation}
where the change of the mutual information $\Delta I$ is split into that of the classical correlation $\Delta J$ and the purely quantum correlation of the initial state of the memory $\delta(B^{(i)}|A^{(i)})$. Equation (\ref{eq_entropy difference_if_2}) is the most important result of our work. If we ignore the well-known contribution of both $\Delta F_S$ and $\Delta S$, the bound of the work extractable is given by two correlations. Even if no classical correlation changes, i.e. $\Delta J =0$, we still find a source of the work given as the discord. This answers the question raised in the beginning; One can extract work from an IHE by using {\em purely quantum mechanical} information contained in the {\em initial} state of the memory, which is expressed as the {\em discord}. It is noted that the quantum discord has already been found in some literatures in the similar context \cite{Zurek03,Oppenheim02}. The discord easily appears once we look for the entropy change of the quantum correlated bipartite system after measuring only one subsystem irrespective of the detailed physical situation.

Note that Eq.~(\ref{eq_entropy difference_if_2}) does not contain the QC mutual information, which differs from the result of Ref.\cite{Sagawa08}. The reason is that we exploit rather a loose inequality (\ref{eq_entropy_total_if}) to derive Eq.~(\ref{eq_work_ext}). It allows us to find more clear expression for the work extracted from purely quantum mechanical information.

Now we show an example of the IHE driven solely by quantum correlation. Let us consider the SZE containing a molecule consisting of two distinct atoms (See Ref.\cite{SW_Kim11} for how thermodynamic processes of the SZE evolves.). Each atom has two fully degenerate internal states designated by $A$ and $B$, which are physically equivalent to two $1/2$ spins. This combined spin system $AB$ plays a role of the memory $M$. We prepare the initial state of $AB$ as the maximally entangled state, namely $\rho_{AB}^{(i)} = \ket{\Psi^+}\bra{\Psi^+}$, with $\ket{\Psi^+} = 1/\sqrt{2}(\ket{\uparrow_A\uparrow_B}+\ket{\downarrow_A \downarrow_B})$. We insert a wall in the middle of the container to separate it into two parts, which completes the stage 1. The SZE is then described by $\rho_S^{(i)} = 1/2\ket{L}\bra{L} + 1/2\ket{R}\bra{R}$, where $\ket{L}$ and $\ket{R}$ denote the state that the molecule is found in the left and the right side, respectively. We regard this the initial state, which is written as
\begin{equation}
\rho^{(i)} = \ket{\Psi^+}\bra{\Psi^+} \otimes 1/2(\ket{L}\bra{L} + \ket{R}\bra{R}) \otimes \rho_R^{can}.
\end{equation}

\begin{figure}
  \includegraphics[width=8.6cm]{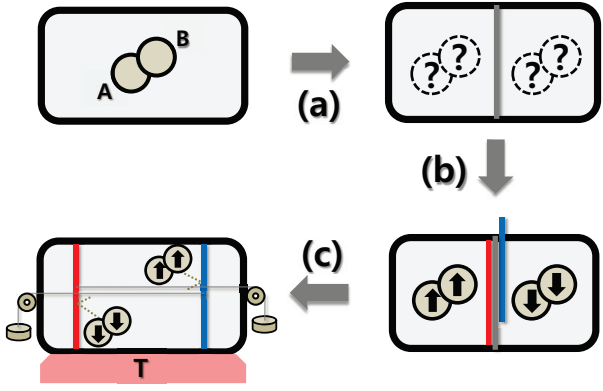}
  \caption{The SZE containing a molecule with two internal states A and B prepared in the Bell's state. (a) A wall, depicted as a vertical gray bar, is inserted to split the box into two parts. The molecule is represented by the dotted circles so as to indicate that at this stage we do not know in which box the molecule is. (b) By applying the unitary operator we have the state given by Eq.~(\ref{eq_ex_rho_(2)}). The SPWs are then inserted so that $W_{\ket{\downarrow}}$ is in the left and $W_{\ket{\uparrow}}$ in the right of the wall, and consequently the wall is removed. (c) A load is attached to each SPW to extract work via an isothermal expansion at a constant temperature T.}
  \label{fig2}
\end{figure}

Next, after detaching the reservoir we apply the unitary operator described by $1/2(\ket{L}\bra{L} \otimes U_L + \ket{R}\bra{R} \otimes U_R)$ with $U_L\ket{\Psi^+} = \ket{\uparrow_A\uparrow_B}$ and $U_R\ket{\Psi^+} = \ket{\downarrow_A\downarrow_B}$ to $\rho^{(i)}$, which generates the coupling between $S$ and $M$. In addition, we perform projection operation onto $A$, which completes the stage 2 of POVM with $\Pi_A^k$ = $\left\lbrace \ket{\uparrow}_A\bra{\uparrow}, \ket{\downarrow}_A\bra{\downarrow}\right\rbrace$. We then obtain
\begin{equation}
\rho^{(2)} = 1/2 ( \ket{\uparrow\uparrow} \bra{\uparrow\uparrow} \otimes \ket{L}\bra{L} + \ket{\downarrow\downarrow}\bra{\downarrow\downarrow} \otimes  \ket{R}\bra{R}),
\label{eq_ex_rho_(2)}
\end{equation}
which implies that the internal state $A$ of the molecule in the left and right side is $\ket{\uparrow}$ and $\ket{\downarrow}$, respectively.

We introduce the most important ingredient of this IHE, a semi-permeable wall (SPW) \cite{Neumann55,Levitin11} denoted as $W_{\ket{\uparrow}(\ket{\downarrow})}$ which prohibits the molecule from passing through it if $A$'s internal state is $\ket{\uparrow}(\ket{\downarrow})$, but becomes transparent if it is $\ket{\uparrow}(\ket{\downarrow})$. In some sense the SPW is similar to a polarizer in optics. Note that the SPW sees only $A$. The SPW's $W_{\ket{\downarrow}}$ and $W_{\ket{\uparrow}}$ are inserted and the wall is removed as shown in Fig.~\ref{fig2}, where the width of all the walls are negligibly small. Now we reattach the reservoir to the engine $S$ and assume that the SPW's are movable. Due to the nature of the SPW's, $W_{\ket{\downarrow}}$ and $W_{\ket{\uparrow}}$ move to the left and to the right, respectively, from which the work of $k_B T \mr{ln}2$ can be extracted via isothermal expansion. This completes the stage 3, a feedback control. The final state is then given as
\begin{equation}
\rho^{(f)} = 1/4 (\ket{\uparrow\uparrow} \bra{\uparrow\uparrow} + \ket{\downarrow\downarrow} \bra{\downarrow\downarrow}) \otimes ( \ket{L}\bra{L} + \ket{R}\bra{R} )\otimes \rho_R^{can}.
\end{equation}
Note that it is guaranteed that the final state of the reservoir satisfies the canonical distribution since the unitary evolution of the total system can describe any thermodynamic processes \cite{Breuer02}.

Where does the work come from? One can easily see $\Delta F_S = 0$ according to $\rho_S^{(i)}=\rho_S^{(f)}$. It is also found that $\rho_X^{(i)}=\rho_X^{(f)}$ due to $\mr{tr}_X[\rho_{AB}^{(i)}] = \mr{tr}_X[\rho_{AB}^{(f)}] = 1/2(\ket{\uparrow}\bra{\uparrow} + \ket{\downarrow}\bra{\downarrow})$ with $X \in \{A,B\}$, implies $\Delta S_A = \Delta S_B = 0$. $\Delta J=0$ is guaranteed from the fact that $\rho_{AB}^{(f)}$ is the post-measurement state of $\rho_{AB}^{(i)}$ on $A$. As far as Eq.~(\ref{eq_work_ext}) is concerned, to avoid violating the second law the work should be originated from the quantum discord. This is confirmed by obtaining $\delta(B^{(i)}|A^{(i)}) = \mr{ln}2$ from Eq.~(\ref{eq_discord}).

The main physics of this engine is summarized as follows. When we focus on the memory during the process, the initial Bell state $\ket{\Psi^+}\bra{\Psi^+}$ with $\ket{\Psi^+}=1/\sqrt{2}(\ket{\uparrow\uparrow}+\ket{\downarrow\downarrow})$ is finally transformed to $1/2(\ket{\uparrow\uparrow}\bra{\uparrow\uparrow} + \ket{\downarrow\downarrow} \bra{\downarrow\downarrow})$. Here the classical correlation, implying if $A$ is $\uparrow$ then $B$ should be $\uparrow$ and vice versa, survives but their quantum superposition, more precisely the entanglement, is broken. The quantumness of this correlation quantified by the discord has been used so as to generate the work.

One might ask why we obtain only $k_BT \mr{ln}2$ instead of $2 k_BT \mr{ln}2$ with the memory of {\em two} bits. The reason is that we have exploited only quantum correlation, the discord. The final state $\rho^{(f)}$ still contains the classical correlation, which can also be used for extracting work by transforming $\rho^{(f)}$ to the fully mixed state, $1/4 (\ket{\uparrow\uparrow}\bra{\uparrow\uparrow} + \ket{\uparrow\downarrow}\bra{\uparrow\downarrow} + \ket{\downarrow\uparrow}\bra{\downarrow\uparrow} + \ket{\downarrow\downarrow} \bra{\downarrow\downarrow})$. We thus extract additional $k_BT \mr{ln}2$ due to $\Delta J = \mr{ln}2$.

Final remark is in order. The work originated from quantum information is not free. The engine considered here is not cyclic in that the memory does not return to the initial Bell state. To recover the initial state one should pay the work equivalent to that obtained during the process, i.e. $k_B T \ln 2$, due to $S(\rho^{(f)})-S(\rho^{(i)}) = \ln 2$.

In summary, we have investigated the bound of the extractable work from the IHE when the correlated memories are taken into account. In addition to the Helmholtz free energy difference and the entropy change of individual memory, the bound contains the quantum mutual information consisting of two parts, the classical correlation and the quantum discord. The quantum discord quantifies the purely quantum mechanical correlation implying that the work can be extracted from purely quantum mechanical information. We confirm it by showing a physical example, a SZE containing a heteronuclear molecule with two atomic internal states initially entangled, where SPW's play a crucial role.

This was supported by the NRF grant funded by the Korea government (MEST) (No.2010-0024644).

\end{document}